\documentclass[]{emulateapj}
   
\begin{document}

\title{Double degenerate mergers as progenitors of high-field magnetic white 
       dwarfs}

\author{Enrique Garc\'\i a--Berro$^{1,2}$,
        Pablo Lor\'en--Aguilar$^{1,2}$,
        Gabriela Aznar--Sigu\'an$^{1,2}$,
        Santiago Torres$^{1,2}$,\\
        Judit Camacho$^{1,2}$,
        Leandro  G.  Althaus$^{3,4}$,
        Alejandro H. C\'orsico$^{3,4}$, 
        Baybars K\"ulebi$^{5,2}$, \&
        Jordi Isern$^{5,2}$}

\affil{$^1$Departament de F\'\i sica Aplicada, 
           Universitat Polit\`ecnica de Catalunya,  
           c/Esteve Terrades, 5,  
           08860 Castelldefels,  
           Spain\\
       $^2$Institute for Space Studies of Catalonia,
           c/Gran Capit\`a 2--4, Edif. Nexus 104,   
           08034  Barcelona, 
	   Spain\\
       $^3$Facultad de Ciencias Astron\'omicas y Geof\'{\i}sicas, 
           Universidad Nacional de La Plata, 
           Paseo del Bosque s/n, 
           (1900) La Plata, 
           Argentina\\
       $^4$Instituto de Astrof\'{\i}sica La Plata, 
           IALP (CCT La Plata), 
           CONICET,
           Argentina\\
       $^5$Institut de Ci\`encies de l'Espai (CSIC), 
           Facultat de Ci\`encies, 
           Campus UAB, 
           Torre C5-parell, 
           08193 Bellaterra, 
           Spain}

\email{garcia@fa.upc.edu}

\begin{abstract}
High-field magnetic  white dwarfs have  been long suspected to  be the
result of stellar mergers. However, the nature of the coalescing stars
and the precise  mechanism that produces the magnetic  field are still
unknown.   Here  we  show  that the  hot,  convective,  differentially
rotating  corona present in  the outer  layers of  the remnant  of the
merger of two  degenerate cores is able to  produce magnetic fields of
the required strength that do  not decay for long timescales.  We also
show,  using  an  state-of-the-art  Monte Carlo  simulator,  that  the
expected number  of high-field magnetic white dwarfs  produced in this
way is consistent with that found in the Solar neighborhood.
\end{abstract}



\section{Introduction}
\label{intro}

The  merger of  two white  dwarfs has  received  considerable interest
during the  last years because  it is thought  to be at the  origin of
several  interesting  astrophysical  phenomena.   In  particular,  the
coalescence of two  white dwarfs is one of  the possible scenarios ---
the so-called  double-degenerate scenario ---  to account for  Type Ia
supernova outbursts  (Webbink 1984; Iben \&  Tutukov 1984).  Moreover,
it is  thought as well that  the merger of two  degenerate cores could
lead to  the formation of  magnetars (King, Pringle  \& Wickramasinghe
2001).   Also, three  hot  and  massive white  dwarfs  members of  the
Galactic  halo could  be the  result of  the coalescence  of  a double
white-dwarf  binary system  (Schmidt et  al.  1992;  Segretain  et al.
1997).  Additionally, hydrogen-deficient  carbon and R Corona Borealis
stars (Izzard et al. 2007; Clayton  et al. 2007; Longland et al. 2011)
are thought to be the consequence  of the merging of two white dwarfs.
Also, the large metal abundances found around some hydrogen-rich white
dwarfs with  dusty disks  around them could  also be explained  by the
merger of a carbon-oxygen and  a helium white dwarf (Garc\'\i a--Berro
et  al.   2007).   Last but  not  least,  the  phase previous  to  the
coalescence of a double white-dwarf close binary system has been shown
to  be  a  powerful  source  of  gravitational  waves  that  would  be
eventually detectable by LISA (Lor\'en--Aguilar et al. 2005).  Here we
show  that the merger  of two  degenerate cores  can also  explain the
presence  of very  high magnetic  fields in  some white  dwarfs  --- a
result previously anticipated by Wickramasinghe \& Ferrario (2000) but
not yet quantitatively proved.

High-field  magnetic white dwarfs  have magnetic  fields in  excess of
10$^6$~G and up to 10$^9$~G  (Schmidt et al. 2003). Surprisingly, very
few belong to a non-interacting binary system (Kawka et al. 2007), and
moreover they are  more massive than average (Silvestri  et al. 2007).
One possibility is that these  white dwarfs descend from single stars,
so the magnetic field is a  fossil of previous evolution (Angel et al.
1981).   However, this  scenario cannot  explain why  these  stars are
massive, and why they are not found in non-interacting binary systems.
Recently,  it  has been  suggested  (Tout  et  al. 2008)  that  strong
magnetic fields  are produced  during a common  envelope episode  in a
close  binary system  in which  one of  the components  is degenerate.
During  this phase,  spiral-in of  the secondary  induces differential
rotation in  the extended convective envelope, resulting  in a stellar
dynamo which produces the magnetic field.  However, the magnetic field
produced in this way does not penetrate in the white dwarf, and decays
rapidly when the common envelope is ejected (Potter \& Tout 2010).

In  this  paper  we   show  that  the  hot,  differentially  rotating,
convective corona  resulting from the  merger of two  degenerate cores
produces  strong magnetic  fields,  which are  confined  to the  outer
layers of the resulting remnant, and  which do not decay for very long
timescales. The  paper is organized as  follows. In Sect.~\ref{dynamo}
we explain  the precise mechanism that produces  the required magnetic
fields, and we show that these fields are confined the outer layers of
the  remnant  of  the coalescence  and  do  not  decay for  very  long
timescales.   Sect.~\ref{MC} is devoted  to analyze  if our  model can
account  for the  number of  high-field magnetic  white dwarfs  in the
Solar  neighborhood,  while   in  Sect.~\ref{conc}  we  summarize  our
findings and we present our conclusions.

\section{The stellar dynamo}
\label{dynamo}

The  merger  of two  degenerate  cores is  the  final  destiny of  the
evolution of  a sizable fraction of  binary systems. Three-dimensional
simulations  of the  merger process  (Guerrero et  al.  2004;  Yoon et
al. 2007; Lor\'en--Aguilar  et al. 2009) indicate that  the remnant of
the coalescence of two white  dwarfs consists of a central white dwarf
which contains all the mass of the primary. On top of it a hot corona,
which  is made  of approximately  half of  the mass  of  the disrupted
secondary, can  be found. Finally, surrounding this  compact remnant a
rapidly rotating  Keplerian disk is formed, containing  nearly all the
mass  of the  secondary which  has not  been incorporated  to  the hot
corona. According  to these calculations little  mass ($\sim 10^{-3}\,
M_{\sun}$) is ejected from the system during the merger. The structure
of the remnant of the coalescence is illustrated in Fig.~1.

The existing simulations show that the temperature gradient in the hot
corona is high, and consequently the corona is convective. We computed
the  boundaries  of  the  convective region  using  the  Schwarzschild
criterion,  and  we  found that  the  inner  and  outer edges  of  the
convective region  are located  at radii $R\approx  0.012\, R_{\sun}$,
and  $R \approx 0.026\,  R_{\sun}$, respectively,  and that  the total
mass   inside  this   region   is  $\sim   0.24   \,  M_{\sun}$   (see
Fig.~1). Moreover, this region rotates differentially, and is prone to
magneto-rotational  instability.  Assuming  energy  equipartition, the
resulting   $\alpha\omega$   dynamo    produces   a   magnetic   field
$B^2/8\pi\approx \rho(\omega R)^2/2$. For  the typical values found in
the simulations of Lor\'en--Aguilar  et al. (2009), the magnetic field
amounts to $B\sim 3.2\times 10^{10}$~G.  Thus, the energy available in
the convective corona is sufficiently large to produce strong magnetic
fields. We note that even in the case in which only 0.1\% of the total
energy  of the  convective shell  is invested  in magnetic  energy the
fields produced  in this  way are  of the order  of $10^7$~G,  a value
typical of high-field magnetic white  dwarfs.  We also note that there
are very few  white dwarfs with magnetic fields  larger than $10^9$~G,
and  that  when these  fields  are  observed,  these are  confined  to
spots on their surfaces.

For this mechanism to be  efficient at producing the observed magnetic
fields, the  dynamo must work for several  convective turnovers before
the energy of the hot corona  is radiated away. The temperature of the
corona is so high that  is preferentially cooled by neutrino emission.
The neutrino  luminosity of the corona is  $L_\nu\sim 4.0\times 10^2\,
L_{\sun}$,  while  the  total  thermal energy  of  the  non-degenerate
material in  the corona is $U\sim 8.8\times  10^{48}$~erg.  Hence, the
convective  shell lasts  for $\tau_{\rm  hot}\sim  1.8\times 10^5$~yr.
The   convective  turnover   timescale   is  $\tau_{\rm   conv}\approx
H_P/v_{\rm  conv}$,  where   $H_P\approx  2.7\times  10^8$~cm  is  the
pressure scale height and $v_{\rm conv}\approx 8.0\times 10^7$ cm/s is
the  convective  velocity.   Thus,  $\tau_{\rm conv}\sim  3.3$~s,  and
during the lifetime of the  hot corona the number of convective cycles
is sufficiently  large. Consequently, the  $\alpha\omega$ mechanism is
able to  produce a strong  magnetic field.  We  also note that  if the
duration of  the convective shell  is substantially smaller  than that
estimated  here,   large  magnetic  fields  can   still  be  produced.
Specifically, even  assuming durations $10^3$ times  smaller than that
previously estimated,  the number of convective cycles  would still be
enough to produce magnetic  fields comparable to those observationally
found in high-field magnetic white  dwarfs.  We thus conclude that the
stellar  dynamos  produced in  the  aftermath  of  the merger  of  two
degenerate cores can produce magnetic fields of the order of $10^7$~G.
From now on we adopt  this value (typical of high-field magnetic white
dwarfs) as a fiducial value for the rest of the calculations.

\begin{figure}[t]
\vspace{7.5cm}
\begin{center} 
\includegraphics{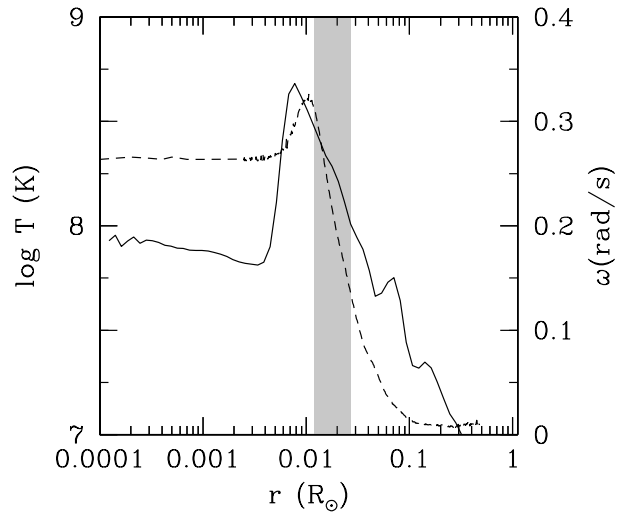} 
\caption{Dynamo  configuration in  a white  dwarf  merger. Temperature
         (solid  line,  left scale)  and  rotational velocity  (dashed
         line, right scale)  profiles of the final remnant  of a white
         dwarf merger as  a function of radius are  shown for the case
         of a  binary system composed of  two stars of  0.6 and $0.8\,
         M_{\sun}$.  In our simulations approximately half of the mass
         of  the  disrupted   secondary  ($\sim  0.3\,  M_{\sun}$)  is
         accreted onto the primary, while the rest of the mass goes to
         form  the Keplerian  disk, which  extends up  to  $\sim 0.3\,
         R_{\sun}$. The temperature of the central white dwarf depends
         on  the initial  temperature of  the coalescing  white dwarfs
         (Yoon et  al. 2007; Lor\'en--Aguilar et  al.  2009). However,
         the temperature  of the hot  corona is remarkably  similar in
         all the simulations, $T\approx 8.7\times 10^8$~K. The central
         spinning  white  dwarf  rotates   as  a  rigid  body  with  a
         rotational  velocity   $\omega\sim  0.26$~s$^{-1}$,  and  the
         corona  rotates  differentially, with  a  peak rotation  rate
         $\omega\sim  0.33$~s$^{-1}$.   These  velocities  arise  from
         energy and  angular momentum conservation,  since little mass
         is ejected  from the system, so the  orbital angular momentum
         of the binary system is  invested in spinning up the remnant,
         while the  available energy is primarily  invested in heating
         the  corona.   The  location  of  the  convective  region  is
         displayed by the shaded area.}
\end{center}
\label{corona} 
\end{figure} 

Once  the magnetic  field is  established we  need to  know if  it can
diffuse  outwards,  to  the  surrounding  disk,  or  inwards,  to  the
degenerate  primary. To  this  end we  solved  the diffusion  equation
(Jackson  1998; Wendell et al. 1987):

\begin{equation}
\frac{\partial\vec{B}}{\partial t}=-\vec{\nabla}\times
\left(\frac{c^2}{4\pi\sigma}\vec{\nabla}\times\vec{B}-
\vec{v}\times\vec{B}\right)
\end{equation}

\noindent being $\sigma$ the  magnetic conductivity, for which we used
the most up-to-date prescription (Cassisi et al. 2007), while the rest
of  the symbols  have their  usual meaning.   We first  integrated the
diffusion  equation   for  the  surrounding   disk  using  cylindrical
coordinates, adopting  the velocity field and  density and temperature
profiles  resulting  from  our  SPH simulations  (Lor\'en--Aguilar  et
al. 2009). The initial condition adopted here is $\vec{B}(0)=\vec{0}$,
while we also imposed the boundary condition $\partial\vec{B}/\partial
t=0$  at  the  outer edge  of  the  disk.   We used  a  Crank-Nicolson
integration scheme with variable  coefficients (Dautray \& Lions 2000)
which  turns  out  to be  stable.  We  found  that the  timescale  for
diffusion  of  the  magnetic  field  across  the  disk  is  $\tau_{\rm
disk}\sim 2.0\times  10^{11}$~yr.  We  did the same  calculation, this
time using  spherical coordinates, to estimate the  penetration of the
magnetic field in the dense, degenerate primary.  The use of spherical
coordinates  is  perfectly justified  because  the  departures of  the
compact primary from sphericity  are very small.  For this calculation
we followed the procedure outlined in Wendell et al.  (1987).  In this
case the  electrical conductivity  is totally dominated  by degenerate
electrons, and  depends on the  adopted temperature of  the isothermal
white dwarf.   If a temperature  $T\approx 7.6\times 10^7$~K  is taken
(Lor\'en--Aguilar et  al.  2009) the diffusion timescale  turns out to
be  $\tau_{\rm  WD}\sim  4.3\times  10^9$~yr.   Since  this  timescale
depends  on the  adopted  initial temperature  we  computed, using  an
up-to-date stellar evolutionary code  (Renedo et al.  2010), a cooling
sequence for a white dwarf of the mass, $1.1\, M_{\sun}$, and chemical
composition of  the remnant, a  carbon-oxygen core.  The time  to cool
from $7.6\times 10^7$~K to a  value typical of field white dwarfs (say
$3.0\times  10^6$~K)  is very  short,  $\tau_{\rm cool}\sim  3.0\times
10^7$~yr. Hence, we  can safely assume that as  the white dwarf cools,
the rapid increase  of the electrical conductivity does  not allow the
magnetic field to penetrate in the degenerate core of the primary, and
remains confined to the surface layers.

Our model  naturally predicts that  the masses of  high-field magnetic
white dwarfs  should be  larger than average  and that they  should be
observed as single white dwarfs, as observationally found --- see, for
instance,  Valyavin \&  Fabrika (1999).  However,  high-field magnetic
white dwarfs  are generally found to be  slow rotators (Wickramasinghe
\& Ferrario  2000).  This apparent  contradiction of the model  can be
easily  solved.  If  the rotation  and magnetic  axes  are misaligned,
magneto-dipole radiation  rapidly spins down the white  dwarf --- see,
however,  Timokhin  (2006)  and  Spitkovsky  (2006),  where  the  time
evolution of  magnetospheres for axisymmetric and  oblique rotators is
described in  detail.  The evolution  of the rotational  velocity when
both axes are misaligned (Benacquista et al. 2003) is:

\begin{equation}
 \dot{\omega} = -\frac{2\mu^2 \omega^3}{3Ic^3}\sin^2\alpha,
\end{equation}

\noindent  where $I$  is the  moment of  inertia of  the  white dwarf,
$\alpha$  is the  angle between  the  magnetic and  rotation axes  and
$\mu=BR_{\rm  WD}^3$.    Thus,  the  spin-down   timescale  is  simply
$\tau_{\rm MDR}=\omega / 2\dot{\omega}$. Adopting the values resulting
from  our SPH  simulations (Lor\'en--Aguilar  et al.  2009)  we obtain
$\tau_{\rm  MDR}\sim  2.4\times10^{8}/\sin^2\alpha$~yr,  for  a  field
strength $B=10^{7}$~G.  Hence, if  both axes are perfectly aligned the
remnant  of the  coalescence will  be a  high-field,  rapidly rotating
magnetic  white dwarf.   On  the  contrary, if  both  axes are  nearly
perpendicular   magneto-dipole   radiation   efficiently  brakes   the
remnant.  Consequently,  the  very  young, hot,  ultramassive,  slowly
rotating  magnetic white  dwarfs 1RXS~J0823.6$-$2525  and PG~1658+441,
which have fields $\sim 3.5  \times 10^6$~G can be accommodated in our
model.  For instance, PG~1658+441 has an effective temperature $T_{\rm
eff}\sim 30\, 000$~K and a  mass $M\sim 1.3\, M_{\sun}$ (Dupuis et al.
2003), which corresponds to a cooling age of $\sim 3.7\times 10^8$~yr,
while  the  time needed  to  brake the  white  dwarf  equals to  $\sim
8.4\times 10^7$~yr if $\sin\alpha=1$  is adopted.  Thus, our model can
account for the  slow rotation rate of PG~1658+441,  provided that the
rotation and magnetic axes are misaligned.  We also note at this point
that given the  small radii --- or, equivalently,  the small radiating
surfaces  --- of  massive white  dwarfs, their  cooling  ages increase
markedly for effective temperatures smaller than $\log T_{\rm eff} \la
4.7$.

On the other hand, rapidly rotating magnetic white dwarfs --- of which
an example is RE~J~0317$-$853,  a very massive ($\sim 1.3\, M_{\sun}$)
white dwarf belonging to a  wide binary system (K\"ulebi et al.  2010)
--- could be  the result  of nearly equal-mass  mergers in  which both
axes are aligned.  In this case  no Keplerian disk is formed, so these
white dwarfs  will not show  infrared excesses.  If, on  the contrary,
the masses of the coalescing  white dwarfs are different, the axes are
not aligned, and the disk is  able to survive for long enough times, a
second-generation   planetary  system   could  be   eventually  formed
(Garc\'\i a--Berro et al.  2007) and tidal disruption and accretion of
minor  planets may  contaminate  the atmosphere  of  the white  dwarf,
resulting  in  magnetic DAZ  or  DAZd  white  dwarfs possibly  showing
infrared  excesses.   Examples of  such  white  dwarfs are  NLTT~10480
(Kawka \& Vennes 2011) and G77$-$5018 (Farihi et al.  2011), for which
a satisfactory explanation is still lacking.

The geometry of  the surface magnetic fields of  white dwarfs has been
investigated over  the years using  spectro-polarimetric observations.
The available wealth of observations  shows that, in almost all cases,
magnetic  white dwarfs  have  fields deviating  strongly from  dipolar
structure.  In particular, it appears  that in most cases the geometry
of  the surface magnetic  field corresponds  to quadrupolar  or higher
multipoles   ---   e.g.,   PG~1031+234   (Schmidt   et   al.    1986),
REJ~0317$-$855 (Ferrario 1997), HE~0241$-$0155 (Reimers et al.  2005),
HE~1045$-$0908 (Euchner et al.   2005), or PG~1015+014 (Euchner et al.
2006).  This means  that the toroidal component of  the magnetic field
must be stable, and that  a poloidal component should also be present.
Our  mechanism can  also qualitatively  reproduce  these observations.
Indeed, in the $\alpha\omega$ mechanism, convection is responsible for
the generation of poloidal fields, whereas rotation is responsible for
the generation of toroidal  fields. Specifically, the energy available
to generate the poloidal field component is $\rho v_{\rm conv}^2/2\sim
4.0\times 10^{20}$~erg,  which is $\sim 10\%$ of  the energy available
to  build the  toroidal component,  $\rho(\omega  R)^2/2\sim 5.5\times
10^{21}$~erg.  Thus,  we expect that  the magnetic field  generated by
the  fast  rotating  convective  shell  will have  both  toroidal  and
poloidal  field components.  Moreover,  it is  well known  that purely
toroidal    fields    are    unstable    due    to    Tayler    (1973)
instability. However, the existence of a small poloidal component is a
sufficient  condition  for   stability.   Actually,  recent  numerical
studies (Braithwaite 2009) show that the energy stored in the poloidal
field component can  be as high as 80\% of  the total magnetic energy,
and  that even poloidal  components $10^{-6}$  times smaller  than the
toroidal one  are enough to warrant stability.   Thus, toroidal fields
generated by the rotating, convective  shell produced in the merger of
two double  degenerates are  stable, and moreover  we expect  that the
magnetic fields generated in this way will have a complex geometry, in
agreement with observations.


\section{Magnetic white dwarfs in the solar neighborhood}
\label{MC}

To  assess  if our  scenario  can  reproduce  the observed  number  of
high-field  magnetic  white  dwarfs  we  have  expanded  an  existing,
state-of-the-art  Monte Carlo  code (Garc\'\i  a--Berro et  al.  1999;
Torres et al.  2002; Garc\'\i a--Berro et al.  2004) designed to study
the Galactic populations of single  white dwarfs to deal with those of
double degenerates.  In our simulations  we assumed that 50\% of stars
belong  to binaries,  and  we  normalized to  the  local disk  density
(Holmberg \&  Flynn 2000).  The  initial primary masses  were obtained
using a standard  initial mass function (Kroupa et  al. 1993), and the
initial  mass  ratios  according  to  a flat  distribution.   Also,  a
constant  star formation  rate and  a  disk age  of 10$^{10}$~yr  were
adopted.  Orbital  separations and eccentricities  were randomly drawn
according  to  a  logarithmic  probability distribution  (Nelemans  et
al. 2001)  and to a thermal distribution  (Heggie 1975), respectively.
For each of  the components of the binary  analytical fits to detailed
stellar  evolutionary tracks were  used (Hurley  et al.   2000), which
provide full coverage from the  main sequence until advances stages of
evolution.  For the  white dwarf stage the most  recent cooling tracks
of Renedo et al. (2010)  were employed.  The orbital evolution of each
binary   was  computed   taking  into   account   circularization  and
synchronization (Hurley et al.  2002).  We also considered mass losses
through stellar  winds.  Specifically,  we assumed that  the evolution
during  the  main sequence  is  conservative,  and  only after  it  we
included  stellar  winds.   The  adopted  mass-loss rate  is  that  of
Kudritzki \& Reimers  (1978) except on the AGB, for  which the rate of
Vassiliadis \& Wood  (1993) was used.  Angular momentum  losses due to
magnetic  braking  and  gravitational  radiation  were  also  included
(Schreiber  et al.   2003; Zorotovic  et al.   2010).  The  Roche lobe
radius  was  computed  using  the  most  commonly  used  approximation
(Eggleton 1983),  and during  the overflow episodes  both rejuvenation
and ageing  were taken  into account (Hurley  et al.  2002).   For the
common  envelope phase  we considered  standard prescriptions  for the
common  envelope  efficiency and  for  the  fraction of  gravitational
binding energy of  the donor available to eject  the envelope --- see,
e.g., Dewi  \& Tauris  (2000).  Specifically, we  adopted $\alpha_{\rm
CE}=0.25$  and  a variable  value  for  the  binding energy  parameter
$\lambda$  (Zorotovic et al.   2010).  With  all these  ingredients we
found that the fraction of merged double degenerate cores in the solar
neighborhood is $\sim 2.9\%$  of the total synthetic population.  This
number includes not only white  dwarf mergers ($\sim 0.3\%$), but also
the coalescence  of a white dwarf  and a giant star  with a degenerate
core  ($\sim 1.1\%$),  and the  merger of  two giants  with degenerate
cores ($\sim 1.5\%$).  In these  two last cases --- namely, the merger
of a white  dwarf and a giant,  and the merger of two  giant stars ---
the  coalescences occur during  the common  envelope phase,  while the
merger of two white dwarfs  is driven by gravitational wave radiation.
Finally, we emphasize that the number of white dwarf mergers we obtain
is in  line with those obtained using  completely different approaches
(Bogomazov \&  Tutukov 2009) ---  see below for a  detailed comparison
with  our  results  for  the  solar  neighborhood  ---  and  that  the
distribution of remnant masses is  nearly flat, in accordance with the
observed distribution of masses of magnetic white dwarfs (Nale{\.z}yty
\& Madej 2004).

\begin{figure}[t]
\vspace{7.5cm}
\begin{center} 
\includegraphics{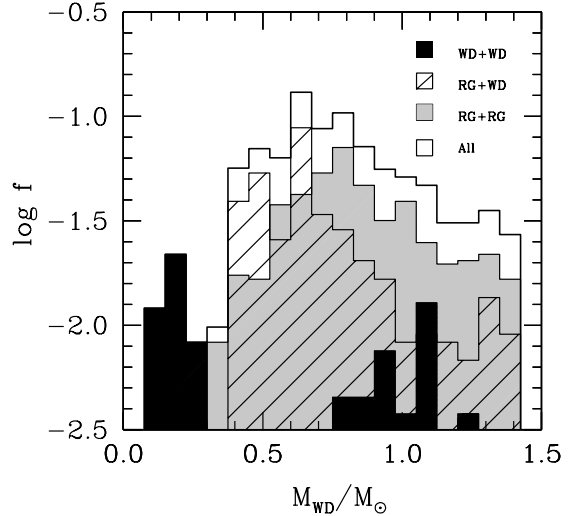} 
\caption{Mass  distribution  of  the  remnants  of  the  mergers.  The
         distribution   shows   the   frequency   of   the   different
         merger channels.  The black histogram shows the masses of the
         remnants of  the mergers of double white  dwarf binaries, the
         dashed  histogram that  of  the mergers  of  a binary  system
         composed  of  a red  giant  and  a  white dwarf,  the  shaded
         histogram that  of the mergers  of two red giants,  while the
         total mass distribution is shown using a solid line.}
\end{center}
\label{population} 
\end{figure} 

To  better illustrate  this  last issue,  Fig.~2  shows the  frequency
distribution of remnant masses of  the different merger channels for a
sample  of  $10^3$ mergers.   We  assumed  that  the remnant  of  each
coalescence  has a mass  $M=M_1+M_2/2$, where  $M_1$ and  $M_2$ stand,
respectively, for  the masses of the primary  and secondary degenerate
cores.   This choice  is in  accordance with  the results  of  the SPH
simulations  of  Lor\'en--Aguilar  et  al.  (2009),  which  show  that
approximately  half of  the secondary  is accreted  onto  the primary,
while the rest  of the mass of the disrupted  secondary forms a debris
region  and  little  mass  is  ejected during  the  merger.   We  also
eliminated   all  those   remnants  that   have  masses   larger  than
Chandrasekhar's limit, while for  the mergers producing a helium white
dwarf we  adopted the procedure of  Hurley et al.  (2002).   As can be
seen, the  total mass distribution  (open histogram) presents  a first
peak for  masses smaller than $\sim 0.4\,  M_{\sun}$, corresponding to
mergers  in which  a  helium  white dwarf  is  produced, then  sharply
increases  for  increasing  remnant  masses  and  afterwards  smoothly
decreases  for masses larger  than $\sim  0.6\, M_{\sun}$.   When this
distribution is sampled for $\sim 26$ objects with masses ranging from
$0.8\,  M_{\sun}$ to $1.4\,  M_{\sun}$ we  usually obtain  fairly flat
distributions,  although   the  scarce  number  of   objects  and  the
subsequent  large  deviations  prevent  a sound  comparison  with  the
observational data.

Within 20  pc of the  Sun there are  122 white dwarfs (Holberg  et al.
2008), and  several of them are  magnetic (Kawka et  al.  2007).  This
sample is 80\%  complete, but subject to poor  statistics. However, it
is useful because for it we  have a reliable determination of the true
incidence  of  magnetism in  white  dwarfs.   Mass determinations  are
available for 121 of them.  In  the local sample there are 14 magnetic
white dwarfs, 8 have magnetic fields in excess of $10^7$~G, and 3 have
masses  larger  than $0.8\,  M_{\sun}$  ---  a  value which  is  $\sim
2.5\sigma$  away from  the average  mass of  field white  dwarfs.  The
selection of this mass cut is somewhat arbitrary but, given the strong
bias introduced by the initial  mass function, we expect that the vast
majority of high-field magnetic  white dwarfs more massive than $0.8\,
M_{\sun}$ would be the result  of stellar mergers.  This is indeed the
case, since  our population synthesis calculations  predict that $\sim
4$ white dwarfs are the  result of double degenerate mergers, and have
masses  larger   than  $0.8\,   M_{\sun}$,  in  good   agreement  with
observations.  Additionally, our simulations predict that the fraction
of white dwarfs more massive than $\sim 0.8\, M_{\sun}$ resulting from
single stellar  evolution is $\sim 10\%$.   Consequently, the expected
number of  massive white  dwarfs in the  local sample should  be $\sim
12$.   Instead,  the local  sample  contains  20,  pointing towards  a
considerable  excess  of massive  white  dwarfs,  which  could be  the
progeny  of mergers.   The rest  of the  population of  magnetic white
dwarfs ($\sim 5$) would be well  the result of the evolution of single
stars (Aznar Cuadrado  et al.  2004). In this  case the magnetic field
could  have been  generated during  the red  giant phase  (Blackman et
al. 2001), or could be a fossil  of the evolution of magnetic Ap or Bp
stars  --- see,  for instance,  Mathys et  al.  (2001)  and references
therein. Note, nevertheless, that  it has been recently suggested that
these  massive main-sequence  stars  are also  the  result of  stellar
mergers (Tutukov \& Fedorova 2010).

The number  of coalescing binaries previously  discussed compares well
with the  results obtained  using very different  population synthesis
codes. For  instance, Bogomazov \& Tutukov (2009),  using a completely
independent  code,  obtain  $\sim  7\pm  1$,  where  we  have  assumed
poissonian errors, white dwarf mergers within 20~pc of the Sun --- see
their Table~1 ---  whereas we obtain $\sim 4\pm  2$.  Thus, within the
sampling uncertainties, both  numbers agree qualitatively, despite the
very   small  number   of   objects  and   the  existing   theoretical
uncertainties, which  include, among others,  the completely different
approaches used  to simulate a representative sample  of binaries, the
ingredients  necessary  to  model  the  Galaxy as  a  whole,  and  the
parameters adopted in the calculation of the common envelope phase.


\section{Summary and conclusions}
\label{conc}

We have shown that the hot, convective, differentially rotating corona
predicted by  detailed three-dimensional simulations of  the merger of
two degenerate  stellar cores  is able to  produce very  high magnetic
fields.  We have also shown that these magnetic fields are confined to
the  outer  layers of  the  remnant of  the  coalescence,  and do  not
propagate neither to the interior of  the white dwarf or to the debris
region.  Our  model naturally predicts that  high-field magnetic white
dwarfs should preferentially have  high masses, and should be observed
as  single white dwarfs,  as observationally  found. Moreover,  if the
rotation  and magnetic  axes are  not aligned  magnetodipole radiation
rapidly  spins down  the  magnetic white  dwarf  in short  timescales.
Thus, high-field  magnetic white  dwarfs may have  a large  variety of
rotation periods.   Moreover, in the case  in which the  masses of the
coalescing white dwarfs are not equal the heavy rotationally supported
disk  can survive for  long periods  of time,  and a  planetary system
could  eventually  form.   Disruption  of  small  planets  could  then
contaminate  the  very  outer  layers  of the  magnetic  white  dwarf,
explaining  the recently  discovered population  of  metallic magnetic
white dwarfs.   If, on the contrary,  the masses of  the merging white
dwarfs are similar the remnant has spherical symmetry and rotates very
rapidly, as  observed in some high-field magnetic  white dwarfs. Also,
the geometry of  the surface magnetic fields can  be well explained by
our model.  Finally, we  have also shown  that the expected  number of
double  degenerate mergers is  roughly consistent  with the  number of
high-field magnetic white dwarfs in the local sample.  In summary, our
calculations  indicate  that  a  sizable fraction  of  all  high-field
magnetic  white  dwarfs  could  be  the result  of  double  degenerate
mergers,  in accordance with  previous suggestions  (Wickramasinghe \&
Ferrario 2000) not hitherto proven.


\acknowledgments

This research  was supported by AGAUR, by  MCINN grants AYA2011--23102
and  AYA08-1839/ESP, by  the European  Union FEDER  funds, by  the ESF
EUROGENESIS  project (grant  EUI2009-04167), by  AGENCIA:  Programa de
Modernizaci\'on  Tecnol\'ogica BID 1728/OC-AR,  and by  PIP 2008-00940
from CONICET.



\begin{thebibliography}{} 
\bibitem[Angel et al.(1981)]{1981ApJS...45..457A} Angel, J.~R.~P., Borra, E.~F., \& Landstreet, J.~D.\ 1981, \apjs, 45, 457
\bibitem[Aznar Cuadrado et al.(2004)]{2004A&A...423.1081A} Aznar Cuadrado, R., Jordan, S., Napiwotzki, R., et al.\ 2004, \aap, 423, 1081 
\bibitem[Benacquista et al.(2003)]{2003ApJ...596L.223B} Benacquista, M., Sedrakian, D.~M., Hairapetyan, M.~V., Shahabasyan, K.~M., \& Sadoyan, A.~A.\ 2003, \apjl, 596, L223
\bibitem[Blackman et al.(2001)]{2001Natur.409..485B} Blackman, E.~G., Frank, A., Markiel, J.~A., Thomas, J.~H., \& Van Horn, H.~M.\ 2001, \nat, 409, 485
\bibitem[Bogomazov \& Tutukov(2009)]{2009ARep...53..214B} Bogomazov, A.~I., \& Tutukov, A.~V.\ 2009, Astronomy Reports, 53, 214 
\bibitem[Braithwaite(2009)]{2009MNRAS.397..763B} Braithwaite, J.\ 2009, \mnras, 397, 763
\bibitem[Cassisi et al.(2007)]{2007ApJ...661.1094C} Cassisi, S., Potekhin, A.~Y., Pietrinferni, A., Catelan, M., \& Salaris, M.\ 2007, \apj, 661, 1094
\bibitem[Clayton et al.(2007)]{2007ApJ...662.1220C} Clayton, G.~C., Geballe, T.~R., Herwig, F., Fryer, C., \& Asplund, M.\ 2007, \apj, 662, 1220
\bibitem[Dautray \& Lions (2000)]{Crank.Nicolson} Dautray, R., \& Lions, J.L.\ 2000, {\sl ``Mathematical Analysis and Numerical Methods for Science and Technology''}, vol. 6 (Heidelberg: Springer Verlag)
\bibitem[Dewi \& Tauris(2000)]{2000A&A...360.1043D} Dewi, J.~D.~M., \& Tauris, T.~M.\ 2000, \aap, 360, 1043
\bibitem[Dupuis et al.(2003)]{2003ApJ...598..486D} Dupuis, J., Chayer, P., Vennes, S., Allard, N.~F., \& H{\'e}brard, G.\ 2003, \apj, 598, 486
\bibitem[Eggleton(1983)]{1983ApJ...268..368E} Eggleton, P.~P.\ 1983, \apj, 268, 368
\bibitem[Euchner et al.(2005)]{2005A&A...442..651E} Euchner, F., Reinsch, K., Jordan, S., Beuermann, K., G\"ansicke, B.~T.\ 2005, \aap, 442, 651 
\bibitem[Euchner et al.(2006)]{2006A&A...451..671E} Euchner, F., Jordan, S., Beuermann, K., Reinsch, K., G\"ansicke, B.~T.\ 2006, \aap, 451, 671 
\bibitem[Farihi et al.(2011)]{2011MNRAS.413.2559F} Farihi, J., Dufour, P., Napiwotzki, R., \& Koester, D.\ 2011, \mnras, 413, 2559
\bibitem[Ferrario et al.(1997)]{1997MNRAS.292..205F} Ferrario, L., Vennes, S., Wickramasinghe, D.~T., Bailey, J.~A., \& Christian, D.~J.\ 1997, \mnras, 292, 205
\bibitem[Garc{\'{\i}}a-Berro et al.(1999)]{1999MNRAS.302..173G} Garc{\'{\i}}a--Berro, E., Torres, S., Isern, J.,\& Burkert, A.\ 1999, \mnras, 302, 173
\bibitem[Garc{\'{\i}}a-Berro et al.(2004)]{2004A&A...418...53G} Garc{\'{\i}}a--Berro, E., Torres, S., Isern, J., \& Burkert, A.\ 2004, \aap, 418, 53
\bibitem[Garc{\'{\i}}a-Berro et al.(2007)]{2007ApJ...661L.179G} Garc{\'{\i}}a--Berro, E., Lor{\'e}n-Aguilar, P., Pedemonte, A.~G., et al.\ 2007, \apjl, 661, L179 
\bibitem[Guerrero et al.(2004)]{2004A&A...413..257G} Guerrero, J., Garc{\'{\i}}a--Berro, E., \& Isern, J.\ 2004, \aap, 413, 257
\bibitem[Heggie(1975)]{1975MNRAS.173..729H} Heggie, D.~C.\ 1975, \mnras, 173, 729
\bibitem[Holberg et al.(2008)]{2008AJ....135.1225H} Holberg, J.~B., Sion, E.~M., Oswalt, T., et al.\ 2008, \aj, 135, 1225
\bibitem[Holmberg \& Flynn(2000)]{2000MNRAS.313..209H} Holmberg, J., \& Flynn, C.\ 2000, \mnras, 313, 209 
\bibitem[Hurley et al.(2000)]{2000MNRAS.315..543H} Hurley, J.~R., Pols, O.~R., \& Tout, C.~A.\ 2000, \mnras, 315, 543
\bibitem[Hurley et al.(2002)]{2002MNRAS.329..897H} Hurley, J.~R., Tout, C.~A., \& Pols, O.~R.\ 2002, \mnras, 329, 897
\bibitem[Iben \& Tutukov (1984)]{1984ApJS...54..335I} Iben, I., Jr., \& Tutukov, A.~V.\ 1984, \apjs, 54, 335
\bibitem[Izzard et al.(2007)]{2007A&A...470..661I} Izzard, R.~G., Jeffery, C.~S., \& Lattanzio, J.\ 2007, \aap, 470, 661
\bibitem[Jackson(1998)]{1998clel.book.....J} Jackson, J.~D.\ 1998, {\sl ``Classical Electrodynamics''}, 3rd Edition, (New York: John Wiley \& Sons)
\bibitem[Kawka et al.(2007)]{2007ApJ...654..499K} Kawka, A., Vennes, S., Schmidt, G.~D., Wickramasinghe, D.~T., \& Koch, R.\ 2007, \apj, 654, 499
\bibitem[Kawka \& Vennes(2011)]{2011A&A...532A...7K} Kawka, A., \& Vennes, S.\ 2011, \aap, 532, A7
\bibitem[King et al. (2001)]{2001MNRAS.320L..45K} King, A.~R., Pringle, J.~E., \& Wickramasinghe, D.~T.\ 2001, \mnras, 320, L45
\bibitem[K{\"u}lebi et al.(2010)]{2010A&A...524A..36K} K{\"u}lebi, B., Jordan, S., Nelan, E., Bastian, U., \& Altmann, M.\ 2010, \aap, 524, A36 
\bibitem[Kudritzki \& Reimers(1978)]{1978A&A....70..227K} Kudritzki, R.~P., \& Reimers, D.\ 1978, \aap, 70, 227
\bibitem[Kroupa et al.(1993)]{1993MNRAS.262..545K} Kroupa, P., Tout, C.~A., \& Gilmore, G.\ 1993, \mnras, 262, 545
\bibitem[Lor{\'e}n-Aguilar et al.(2005)]{2005MNRAS.356..627L} Lor{\'e}n-Aguilar, P., Guerrero, J., Isern, J., Lobo, J.~A., \& Garc{\'{\i}}a--Berro, E.\ 2005, \mnras, 356, 627
\bibitem[Lor{\'e}n-Aguilar et al.(2009)]{2009A&A...500.1193L} Lor{\'e}n--Aguilar, P., Isern, J., \& Garc{\'{\i}}a--Berro, E.\ 2009, \aap, 500, 1193
\bibitem[Longland et al.(2011)]{2011ApJ...737L..34L} Longland, R., Lor{\'e}n-Aguilar, P., Jos{\'e}, J., et al.\ 2011, \apjl, 737, L34
\bibitem[Mathys et al.(2001)]{2001ASPC..248.....M} Mathys, G., Solanki, S.~K., \& Wickramasinghe, D.~T.\ 2001, {\sl ``Magnetic Fields Across the Hertzsprung-Russell Diagram''}, ASP Conf. Proc. vol. 248 (San Francisco: Astron. Soc. of the Pacific)
\bibitem[Nale{\.z}yty \& Madej(2004)]{2004A&A...420..507N} Nale{\.z}yty, M., \& Madej, J.\ 2004, \aap, 420, 507
\bibitem[Nelemans et al.(2001)]{2001A&A...365..491N} Nelemans, G., Yungelson, L.~R., Portegies Zwart, S.~F., \& Verbunt, F.\ 2001, \aap, 365, 491 
\bibitem[Potter \& Tout(2010)]{2010MNRAS.402.1072P} Potter, A.~T., \& Tout, C.~A.\ 2010, \mnras, 402, 1072
\bibitem[Reimers et al.(2004)]{2004A&A...414.1105R} Reimers, D., Jordan, S., \& Christlieb, N.\ 2004, \aap, 414, 1105
\bibitem[Renedo et al.(2010)]{2010ApJ...717..183R} Renedo, I., Althaus, L.~G., Miller Bertolami, M.~M., et al.\ 2010, \apj, 717, 183 
\bibitem[Schmidt et al.(1986)]{1986ApJ...309..218S} Schmidt, G.~D., West, S.~C., Liebert, J., Green, R.~F., \& Stockman, H.~S.\ 1986, \apj, 309, 218
\bibitem[Schmidt et al. (1992)]{1992ApJ...394..603S} Schmidt, G.~D., Bergeron, P., Liebert, J., \& Saffer, R.~A.\ 1992, \apj, 394, 603
\bibitem[Schmidt et al.(2003)]{2003ApJ...595.1101S} Schmidt, G.~D., Harris, H.~C., Liebert, J., et al.\ 2003, \apj, 595, 1101
\bibitem[Schreiber \& G\"ansicke(2003)]{2003A&A...406..305S} Schreiber, M.~R., \& G\"ansicke, B.~T.\ 2003, \aap, 406, 305 
\bibitem[Segretain et al.(1997)]{1997ApJ...481..355S} Segretain, L., Chabrier, G., \& Mochkovitch, R.\ 1997, \apj, 481, 355
\bibitem[Silvestri et al.(2007)]{2007AJ....134..741S} Silvestri, N.~M., Lemagie, M.~P., Hawley, S.~L., et al.\ 2007, \aj, 134, 741
\bibitem[Spitkovsky(2006)]{2006ApJ...648L..51S} Spitkovsky, A.\ 2006, \apjl, 648, L51
\bibitem[Tayler(1973)]{1973MNRAS.161..365T} Tayler, R.~J.\ 1973, \mnras, 161, 365 
\bibitem[Timokhin(2006)]{2006MNRAS.368.1055T} Timokhin, A.~N.\ 2006, \mnras, 368, 1055
\bibitem[Torres et al.(2002)]{2002MNRAS.336..971T} Torres, S., Garc{\'{\i}}a--Berro, E., Burkert, A., \& Isern, J.\ 2002, \mnras, 336, 971
\bibitem[Tout et al.(2008)]{2008MNRAS.387..897T} Tout, C.~A., Wickramasinghe, D.~T., Liebert, J., Ferrario, L., \& Pringle, J.~E.\ 2008, \mnras, 387, 897
\bibitem[Tutukov \& Fedorova(2010)]{2010ARep...54..156T} Tutukov, A.~V., \& Fedorova, A.~V.\ 2010, Astronomy Reports, 54, 156 
\bibitem[Valyavin \& Fabrika(1999)]{1999ASPC..169..206V} Valyavin, G., \& Fabrika, S.\ 1999, in {\sl ``Proc. of the 11th European Workshop on White Dwarfs''}, Eds.: S.E. Solheim \& E.G. Meistas, ASP Conf. Proc. vol. 169, 206 (San Francisco: Astron. Soc. of the Pacific)
\bibitem[Vassiliadis \& Wood(1993)]{1993ApJ...413..641V} Vassiliadis, E., \& Wood, P.~R.\ 1993, \apj, 413, 641 
\bibitem[Webbink (1984)]{1984ApJ...277..355W} Webbink, R.~F.\ 1984, \apj, 277, 355
\bibitem[Wendell et al. (1987)]{1987ApJ...313..284W} Wendell, C.~E., van Horn, H.~M., \& Sargent, D.\ 1987, \apj, 313, 284 
\bibitem[Wickramasinghe \& Ferrario(2000)]{2000PASP..112..873W} Wickramasinghe, D.~T., \& Ferrario, L.\ 2000, \pasp, 112, 873
\bibitem[Yoon et al.(2007)]{2007MNRAS.380..933Y} Yoon, S.-C., Podsiadlowski, P., \& Rosswog, S.\ 2007, \mnras, 380, 933
\bibitem[Zorotovic et al.(2010)]{2010A&A...520A..86Z} Zorotovic, M., Schreiber, M.~R., G{\"a}nsicke, B.~T., \& Nebot G{\'o}mez-Mor{\'a}n, A.\ 2010, \aap, 520, A86 
\end{thebibliography}
\end{document}